\documentclass[prl,aps,twocolumn,amssymb,amsmath,superscriptaddress,showpacs,aps,floatfix]{revtex4}
\newcommand{\be}{\begin{equation}}
\newcommand{\ee}{\end{equation}}
\newcommand{\bea}{\begin{eqnarray}}
\newcommand{\eea}{\end{eqnarray}}
\usepackage{graphicx}
\usepackage{epsfig}
\usepackage{bbold}

\newcommand{\ah}{\hat{a}}
\newcommand{\ahd}{\hat{a}^\dagger}

\newcommand{\chd}{\hat{c}^\dagger}

\newcommand{\Sh}{\hat{S}}

\newcommand{\Uh}{\hat{U}}
\newcommand{\Uhd}{\hat{U}^\dagger}
\newcommand{\eq}[1]{(\ref{eq:#1})}
\newcommand{\eqname}[1]{\label{eq:#1}}

\begin{document}

\title{Back-action effects in an all-optical model of dynamical Casimir emission}
\author{I. Carusotto}
\email{carusott@science.unitn.it}

\affiliation{INO-CNR BEC Center and Dipartimento di Fisica, Universit\`a di Trento, I-38123 Povo, Italy}

\author{S. De Liberato}
\affiliation{Laboratoire Mat\'eriaux et Ph\'enom\`enes Quantiques,
Universit\'e Paris Diderot-Paris 7 and CNRS, UMR 7162, 75205 Paris
Cedex 13, France}

\author{D. Gerace}
\affiliation{CNISM-UdR Pavia and Dipartimento di Fisica ``A. Volta'',  Universit\`a di Pavia, 27100 Pavia, Italy}

\author{C. Ciuti}
\affiliation{Laboratoire Mat\'eriaux et Ph\'enom\`enes Quantiques,
Universit\'e Paris Diderot-Paris 7 and CNRS, UMR 7162, 75205 Paris
Cedex 13, France}
\date{\today}

\begin{abstract}
We report a theoretical study of the optical properties of a three-level emitter embedded in an optical cavity including the non-rotating wave terms of the light-matter interaction Hamiltonian. Rabi oscillations induced by a continuous wave drive laser are responsible for a periodic time-modulation of the effective cavity resonance, which results in a significant dynamical Casimir emission. A clear signature of the back-action effect of the dynamical Casimir emission onto the drive laser is visible as a sizable suppression of its absorption.
\end{abstract}


\maketitle

One of the most intriguing predictions of modern quantum field theory is the possibility of converting the zero-point fluctuations of a quantum field into real particles when the boundary conditions of the field are varied in time at a fast enough rate. The most celebrated example of such an effect is the emission of electromagnetic radiation from a moving but electrically neutral metallic mirror, the so called dynamical Casimir effect (DCE)~\cite{DCE}. 

So far, experimental observation of the DCE has been hindered by the extremely weak intensity of the emission in realistic configurations~\cite{Lambrecht_rev}, but theoretical understanding of the underlying physics can be considered as almost satisfactory. The next theoretical challenge consists in combining the quantum field theory with a microscopic description of the mirror motion, including the mechanical back-action effect of the quantum field onto the mirrors: pioneering theoretical works have anticipated the appearance of a very weak friction force to compensate the radiated energy~\cite{Kardar}, but to the best of our knowledge no experimental investigation has been reported yet.

New perspectives in the study of the dynamical Casimir effect were opened by the realization that a sizable DCE can be obtained with a suitable time-modulation of the effective optical length of the cavity: many proposals to very the optical length have been put forward, from photo-generation of a highly reflecting plasma in the mirrors~\cite{Yablo,Braggio}, to time-modulations of the refractive index of the cavity material~\cite{DCrefr}, to coupling the cavity mode to an emitter with time-dependent properties~\cite{simone,atomi,qubit}. First experimental evidences of a DCE have recently appeared using a superconducting circuit terminated by a SQUID~\cite{delsing}; as theoretically proposed in~\cite{johansson}, the effective electric length of the circuit is modulated by driving the SQUID with a time-dependent magnetic flux.

\begin{figure}[htbp]
\parbox[c]{0.4\columnwidth}{\includegraphics[width=0.4\columnwidth,clip]{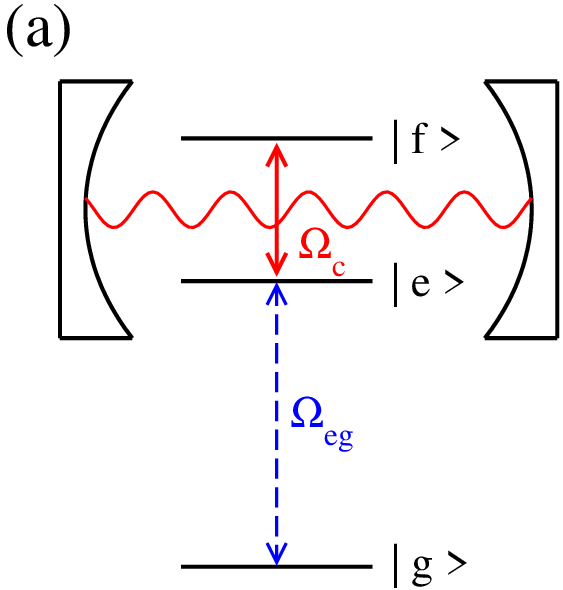}\vspace*{1.cm}}
\hspace{0.02\columnwidth}
\parbox[c]{0.55\columnwidth}{\includegraphics[width=0.55\columnwidth,clip]{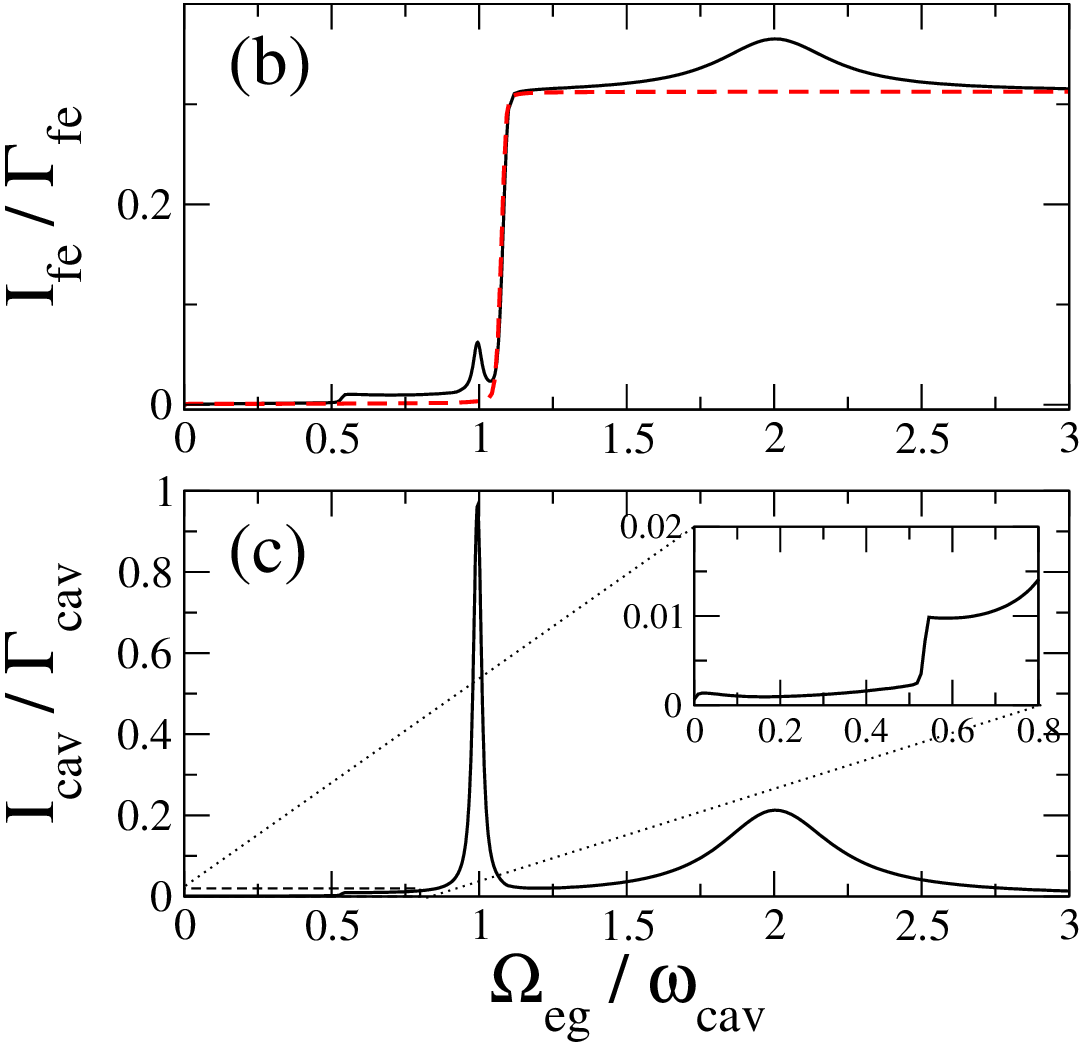}}
\caption{(Color online) Left (a) panel: sketch of the ladder level configuration in the emitter coupled to the cavity and of the optical fields driving the transitions.
Right panels: total emission intensities $I_{\rm cav}$ and $I_{fe}$ as a function of the Rabi frequency $\Omega_{eg}$ of the drive laser (solid lines). Red dashed line in (b): same curve in the vanishing emitter-cavity coupling case $\Omega_{\rm cav}=0$. Within the RWA both $I_{{\rm cav},ge}$ would be exactly zero. 
The chosen parameters $\omega_e-\omega_g=\omega_L$, $\omega_f-\omega_e=\omega_{\rm cav}$, $\Omega_{\rm cav}/\omega_{\rm cav}=0.1$, $\Gamma_{eg}/\omega_{\rm cav}=0.01$, $\Gamma_{\rm cav}=\Gamma_{fe}=10^{-3}\,\omega_{\rm cav}$ correspond to a strong (but not ultra-strong) coupling regime. 
\label{fig:scheme_intensity}}
\end{figure}

In this Letter we consider a model of a cavity whose effective optical length is modulated in time in an all-optical way by means of a drive laser beam that induces Rabi oscillations in a three-level emitter in a ladder configuration strongly coupled to an optical cavity, as sketched in Fig.\ref{fig:scheme_intensity}(a). Such a configuration has been recently realized in the experiment of~\cite{huber} using a combination of inter-band and inter-subband transitions in a semiconductor device and a theoretical investigation of its optical properties at the level of the rotating wave approximation (RWA) has appeared in~\cite{ridolfo}. Of course, the same model is applicable to other physical realizations, such as atoms in microwave cavities~\cite{haroche}, Josephson qubits in superconducting circuits~\cite{circuits}, or quantum dots in photonic crystal cavities~\cite{imam}.

Here we go beyond the RWA and perform a complete quantum optical calculation of the system dynamics including all the relevant non-RWA terms in the light-matter interaction Hamiltonian as well as in the dissipation superoperators. In particular, we anticipate the appearance of an appreciable quantum emission by dynamical Casimir effect. More importantly, the optical nature of the cavity modulation allows to read out the back-action effect of the quantum emission from the absorption that the drive laser experiences upon interacting with the emitter. This effect is the optical analog of the friction force experienced by the moving mirrors in standard DCE configurations~\cite{Lambrecht_rev}, but its actual measurement is expected to take great advantage of the extreme precision of optical techniques over mechanical ones.

The internal dynamics of the emitter-cavity system is described by the model Hamiltonian
\begin{multline}
H=\hbar \omega_{\rm cav} \ahd \ah + 
\hbar \omega_g |g\rangle\langle g| +
\hbar \omega_e |e\rangle\langle e| +
\hbar \omega_f |f\rangle\langle f| + \\
+ {\hbar\Omega_{eg}}\,e^{-i\omega_L t} |e\rangle\langle g| + {\hbar\Omega_{eg}}\,e^{i\omega_L t} \,|g\rangle\langle e|  + \\
+ {\hbar\Omega_{\rm cav}} \, \left[ |f\rangle\langle e| + |e\rangle\langle f| \right] \left[\ahd + \ah\right],
\eqname{hamiltonian}
\end{multline}
where $\hbar\omega_{g,e,f}$ are the energies of the $g,e,f$ atomic levels, respectively.
The $g\leftrightarrow e$ transition is optically driven by a coherent laser of frequency $\omega_L$ and (real and positive) Rabi frequency $\Omega_{eg}$. The standard RWA has been performed on this transition under the assumption that its frequency $\omega_e-\omega_g$ is much higher than all other frequency scales.
A single cavity mode of frequency $\omega_{\rm cav}$ and destruction (creation) operator  $\ah$ ($\ahd$) is considered, which is strongly coupled to the $e\rightarrow f$ transition with a (real and positive) vacuum Rabi frequency $\Omega_{\rm cav}$. In order to correctly describe the DCE, all terms of this coupling have to be taken into account, including the anti-RWA ones where a cavity photon is emitted while the atom climbs from the $e$ to the $f$ state and viceversa. In the following, we shall restrict our attention to the resonant case with $\omega_L=\omega_e-\omega_g$ and $\omega_{\rm cav}=\omega_f-\omega_e$. As required by the strong light-matter coupling regime, the light-matter coupling $\Omega_{{\rm cav}}$ is assumed to be much larger than all decay rates.

\begin{figure}[htbp]
\hspace{0.0\columnwidth}
\parbox[c]{0.525\columnwidth}{\includegraphics[width=0.525\columnwidth,clip]{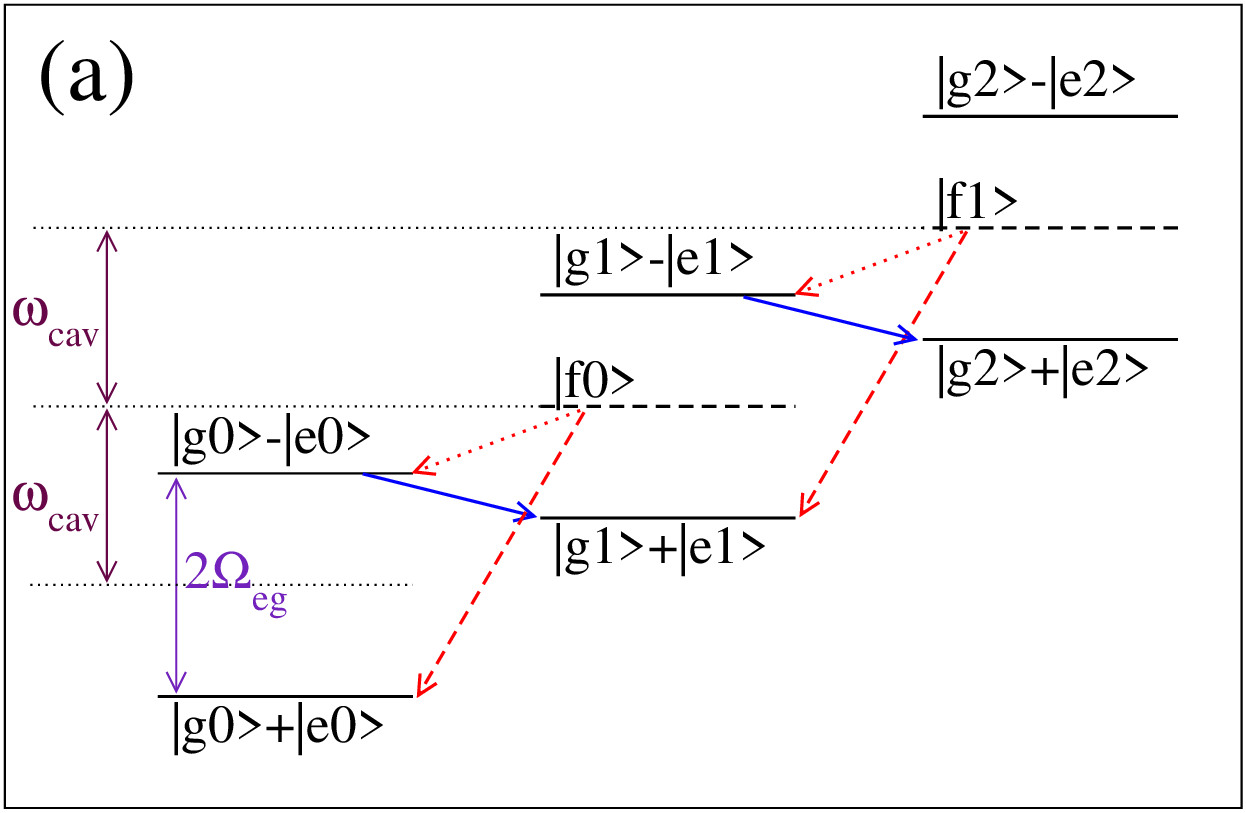}}
\hspace{0.01\columnwidth}
\parbox[c]{0.425\columnwidth}{\includegraphics[width=0.425\columnwidth,clip]{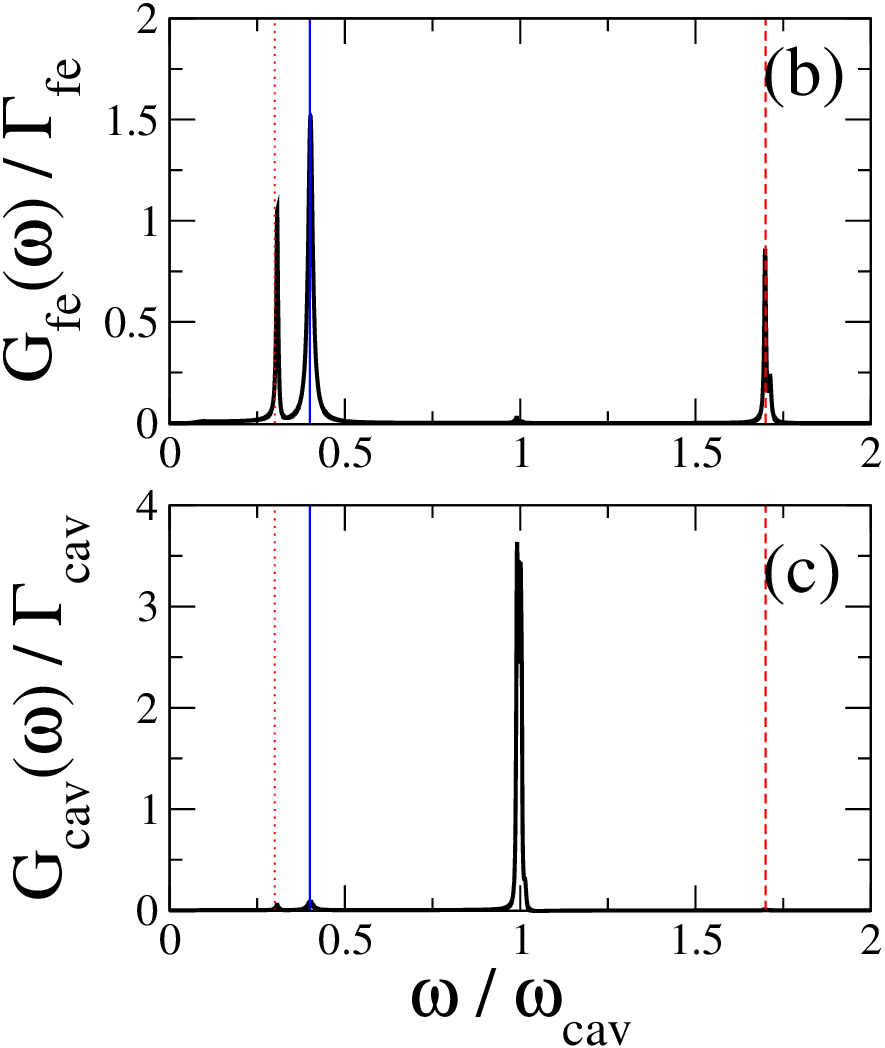}}
\vspace{0.25cm}
\hspace{0.0\columnwidth}
\parbox[c]{0.525\columnwidth}{\includegraphics[width=0.525\columnwidth,clip]{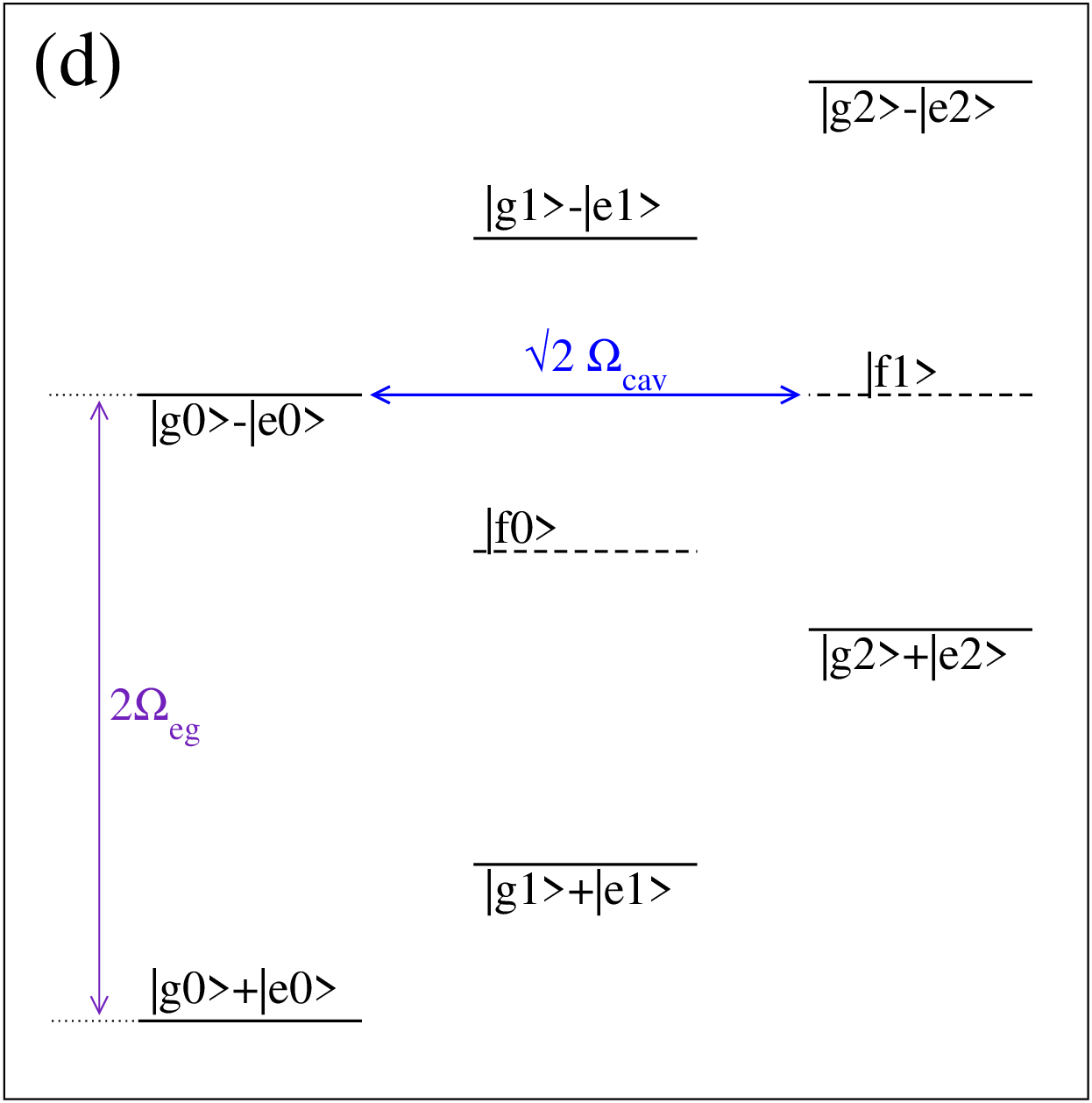}}
\hspace{0.01\columnwidth}
\parbox[c]{0.425\columnwidth}{\includegraphics[width=0.425\columnwidth,clip]{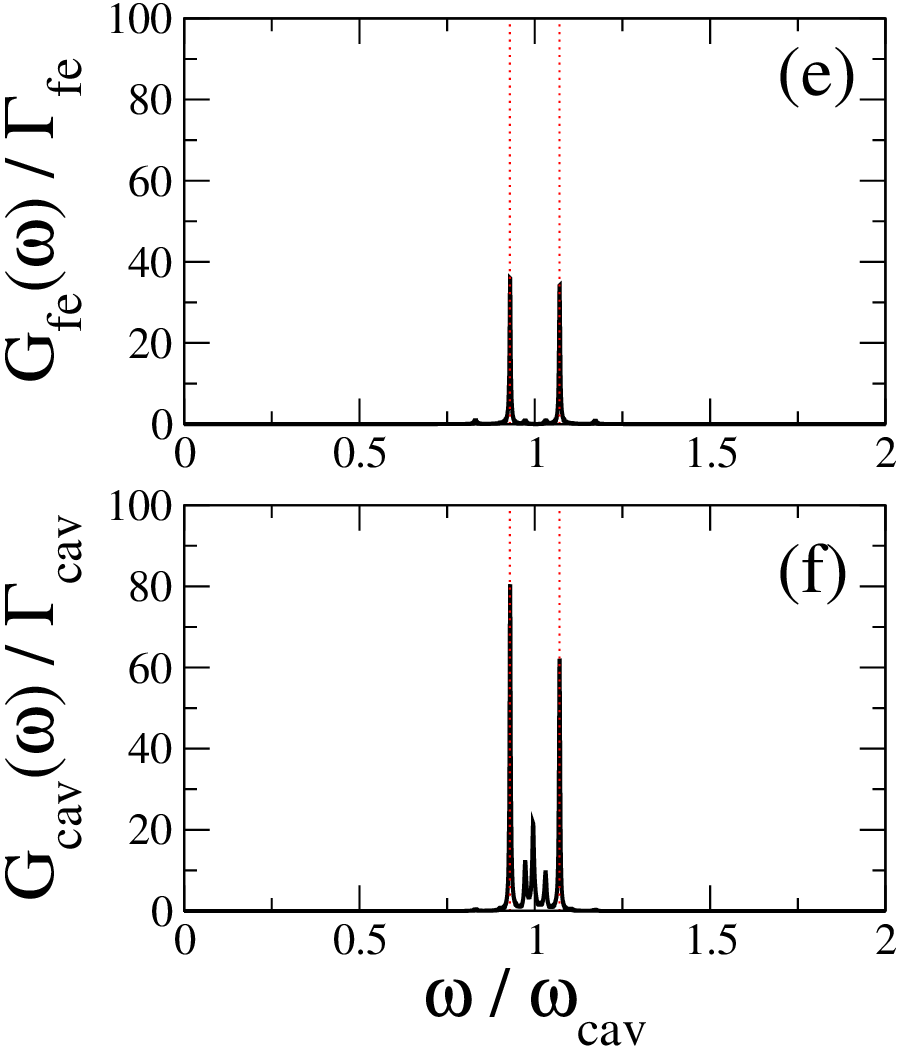}}
\caption{(Color online) Left (a,d) panels: sketch of the dressed levels of the optically driven emitter as predicted by the Hamiltonian \eq{hamiltonian_dressed}. The labels indicate the dominant contribution of each eigenstate in the weak atom-cavity coupling $\Omega_{\rm cav}$ limit. Right panels: spectra of the spontaneous emission on the $f\leftrightarrow e$ transition (b,e) and of the cavity emission (c,f).
Upper and lower group of panels (a-c) and (d-f) refer to the $\Omega_{eg}/\omega_{\rm cav}=0.7, \,2$ cases, respectively.
Same system parameters as in Fig.\ref{fig:scheme_intensity}.
\label{fig:levels_spectra}}
\end{figure}

 Thanks to the RWA assumption on the $g\leftrightarrow e$ transition, we can apply the unitary rotation operator $R(t)=e^{-i\omega_Lt\,|g\rangle\langle g|}$ and move to a rotating frame where the Hamiltonian has a time-independent form
\begin{multline}
H=\hbar \omega_{\rm cav} \ahd \ah + 
\hbar (\omega_g+\omega_L) |g\rangle\langle g| +
\hbar \omega_e |e\rangle\langle e| +
\hbar \omega_f |f\rangle\langle f| + \\
+ {\hbar\Omega_{eg}}\left[|e\rangle\langle g| + |g\rangle\langle e|\right]  + \\
+ {\hbar\Omega_{\rm cav}} \, \left[ |f\rangle\langle e| + |e\rangle\langle f| \right] \left[\ahd + \ah\right].
\eqname{hamiltonian_dressed}
\end{multline}
The structure of the resulting eigenstates of the cavity-emitter system optically dressed by the drive laser is shown in Fig.\ref{fig:levels_spectra}(a,d) for different values of $\Omega_{eg}$. The labels of the eigenstates refer to the dominating component in the weak $\Omega_{\rm cav}$ limit, in which the energy of the $|gn\rangle\pm |en\rangle$ and $|fn\rangle$ eigenstates are $n\omega_{\rm cav}\mp\Omega_{eg}$ and $(n+1)\omega_{\rm cav}$ respectively. In the shorthand $|jn\rangle$, $n$ and $j=\{g,e,f\}$ respectively indicate the number of cavity photons and the state of the emitter.

In addition to the cw driving laser, the system is coupled to the external world via spontaneous emission processes on the $e\rightarrow g$ and $f\leftrightarrow e$ transitions (the $f\rightarrow g$ transition is assumed to be forbidden), as well as via direct light emission from the cavity through the non-perfectly reflecting mirrors.
Such dissipation processes are included in the model at the level of the master equation for the density matrix $\rho$,
\begin{equation}
\frac{d\rho}{dt}=-\frac{i}{\hbar}[H,\rho]+\mathcal{L}_{eg}[\rho]+\mathcal{L}_{fe}[\rho]+\mathcal{L}_{\rm cav}[\rho].
\eqname{master}
\end{equation}
Thanks to the large value of $\omega_e-\omega_g$, the standard RWA approximation can be performed on the $e \rightarrow g$ spontaneous emission terms, which leads~\cite{QuantumNoise} to a superoperator which has the usual Lindblad form, $\mathcal{L}_{eg}[\rho]=\frac{1}{2}\Gamma_{eg}[2\sigma^-_{eg}\rho\sigma^+_{eg}-\sigma^+_{eg}\sigma^-_{eg}\rho-\rho\sigma^+_{eg}\sigma^-_{eg}]$ in both the stationary and rotating frames, with $\sigma^+_{eg}=|e\rangle\langle g|$ and $\sigma^-_{eg}=(\sigma_{eg}^+)^\dagger$.
Nonetheless, it is worth reminding that energy conservation shows peculiar features when seen in the rotating frame of the dressed Hamiltonian \eq{hamiltonian_dressed}: as the emission spectrum on the $e\rightarrow g$ transition is shifted downwards by $\omega_L$, the sidebands on the red side of $\omega_L=\omega_e-\omega_g$ have a negative energy in the rotating frame; on the dressed level scheme of Fig.\ref{fig:levels_spectra}(a,d), these transitions appear in fact as climbing up the energy ladder.

Consistency of the theoretical model requires including the anti-RWA terms for the spontaneous emission on both the $f\leftrightarrow e$ transition and in the cavity emission. This requires some additional care as a number of counter-intuitive optical processes may occur where, e.g., the atom goes from the $e$ to the $f$ state while emitting a photon by spontaneous emission. For a sufficiently weak coupling to the baths~\cite{petruccione}, the dissipation superoperators can be written in the temporally local form
\begin{equation}
\mathcal{L}_{j}[\rho]=\Gamma_j\,\{\hat{U}_j\rho \hat{S}_j + \Sh_j\rho \Uhd_j -\Sh_j \Uh_j \rho - \rho \Uhd_j \Sh_j\},
\eqname{superoperators}
\end{equation}
where the system-bath interaction operators for $j=\{fe,{\rm cav}\}$ have the form $\Sh_{fe}=|e\rangle\langle f|+|f\rangle\langle e|$ and $\Sh_{\rm cav}=\ahd+\ah$, respectively. 
Energy conservation in the decay process is implemented by the integral operators 
\begin{equation}
\Uh_j=\int_0^\infty\!d\tau\,v_j(\tau)\,e^{-iH\tau}\,\Sh_j\,e^{iH\tau},
\end{equation}
whose definition involves the Fourier transform of the frequency-dependent density of states of the baths,
$v_j(\tau)=(2\pi)^{-1} \int_{-\infty}^\infty\!d\omega\,e^{-i\omega\tau}\,v_j(\omega)$.
As a result, the decay from the $|\textrm{in}\rangle$ to the $|\textrm{fin}\rangle$ eigenstate of the dressed system occurs at a rate $\Gamma_j\,|\langle \textrm{fin}|S_j|\textrm{in}\rangle|^2\,v(\omega_{\textrm{in}}-\omega_{\textrm{fin}})$.
In the present paper, all baths are assumed to be zero temperature ones, so that energy can only dissipated from the system into the bath, which imposes $v(\omega)=0$ for $\omega<0$.
For the numerical calculations shown in the figures, we have considered model forms of $v_j(\omega)$ such that $v_j(\omega)=0$ for either $\omega<0$ or $\omega>\omega_{\rm max}$, with a UV cut-off $\omega_{\rm max}$ chosen to be much higher than all other energy scales of the problem. In the intermediate region $0<\omega<\omega_{\rm max}$, $v_j(\omega)$ is taken to be flat and equal to $1$ exception made for a smoothening of the edges. We have checked that all physical conclusions do not critically depend on the specific choice of the UV cut-off $\omega_{\rm max}$ and of the details of the smoothening~\footnote{In the numerics, the $v_j(\omega)$'s have been smoothened with a Gaussian tail of width $\Delta \omega=0.025\omega_{\rm cav}$ past the extrema at $\omega_{\rm edge}=0.1\, \omega_{\rm cav}$ and $\omega_{\rm max}=21\,\omega_{\rm cav}$.}.

The expectation value of all one-time operators of the system in the stationary state are immediately obtained from the steady-state solution $\rho_{\rm ss}$ of the master equation \eq{master}, in particular the full photon number distribution in the cavity. 
However, as it was pointed out in~\cite{bastard}, this is not enough to determine the actual light emission from the system: to distinguish real radiation from the virtual photons that are present even in the ground state of the cavity, one has in fact to evaluate the full spectral distribution of the emitted radiation~\cite{QuantumNoise}, that is the Fourier transform of the two-time correlation function\footnote{The two-time averages are defined as $\langle A(t+\tau)B(t)\rangle=\textrm{Tr}[A\rho'(t+\tau)]$, where $\rho'(t+\tau)$ is the solution of  \eq{master} starting from the initial condition $\rho'(t)=B\rho(t)$. For $\tau<0$, one uses 
formula $\langle A(t+\tau) B(t) \rangle=\langle B^\dagger(t) A^\dagger(t+\tau) \rangle^*$.}
\begin{equation}
G_{j}(\omega)=\Gamma_{j}\,v_{\rm cav}(\omega)\,\int_{-\infty}^\infty\!\!\!\!\!d\tau\,e^{i\omega\tau}\left\langle 
S_{j}(t+\tau)\,S_{j}(t)\right\rangle
\eqname{G}
\end{equation}
for $j=\{{\rm cav},fe\}$, from which the total emission intensities $I_j$ are obtained after integration over $\omega$.
Plots of $I_{{\rm cav},fe}$ as a function of $\Omega_{eg}$ are shown in Fig.\ref{fig:scheme_intensity}(b,c). 

Both emissions have a negligible value~\footnote{The non-vanishing values of $I_{{\rm cav},fe}$ down to $\Omega_{eg}=0$ are an artifact of the weak-coupling approximation to write the master equation in the temporally local form \eq{master}.} for low Rabi frequencies $\Omega_{eg}<\omega_{\rm cav}/2$. The first threshold occurs at $\Omega_{eg}=\omega_{\rm cav}/2$. A second, more pronounced threshold is apparent at $\Omega_{eg}=\omega_{\rm cav}$ and is followed by additional structure in $I_{\rm cav}$ for larger $\Omega_{eg}$'s, in particular two peaks at $\Omega_{eg}=\omega_{\rm cav}$ and $2\omega_{\rm cav}$. 
When the emitter is not coupled to the cavity ($\Omega_{\rm cav}=0$, red dashed line in Fig.\ref{fig:scheme_intensity}(b)), the  lower threshold and the two peaks disappear, while the upper threshold is almost unaffected. On the other hand, the emission completely vanishes if the anti-RWA terms are not included.

A physical explanation of these numerical observations is obtained by looking at the level schemes of the optically dressed system that are shown in Fig.\ref{fig:levels_spectra}(a,d).
For $\Omega_{eg}<\omega_{\rm cav}/2$, the dynamics of the system is limited to the subspace spanned by the two lowest energy eigenstates $|g,0\rangle\pm|e,0\rangle$ at energies $\mp\hbar\Omega_{eg}$.  For $\Omega_{eg}>\omega_{\rm cav}/2$, the energy of the $|gn\rangle-|en\rangle$ dressed states starts exceeding the energy of the $|g(n+1)\rangle+|e(n+1)\rangle$ state, which activates a novel family of decay processes indicated by the blue solid arrows in Fig.\ref{fig:levels_spectra}(a). These decays occur via spontaneous emission on the $e\rightarrow f$ transition, the matrix element being provided by the weak admixture (proportional to $\Omega_{\rm cav}/\Omega_{eg}$) of the neighboring $|fn\rangle$ state into the final $|g(n+1)\rangle+|e(n+1)\rangle$ state by the emitter-cavity coupling $\Omega_{\rm cav}$.
As a result, population is transferred to the upper manifolds and significant emission intensities $I_{{\rm cav},fe}$ appear as the dressed system goes back to the lowest manifold.
The second, stronger threshold that is visible in Fig.\ref{fig:scheme_intensity}(b,c) at $\Omega_{eg}=\omega_{\rm cav}$ is due to an additional $e\rightarrow f$ spontaneous decay channel that opens up as soon as the $|gn\rangle+|en\rangle$ state exceeds in energy the $|fn\rangle$ state. 

This interpretation of the thresholds is confirmed by the plots of the frequency-resolved emission spectra $G_{\rm cav}(\omega)$ and $G_{fe}(\omega)$ shown in Fig.\ref{fig:levels_spectra}(b,c). Note that, in contrast to the previous discussion of the $e\rightarrow g$ emission, the frequency $\omega$ in $G_{{\rm cav},fe}(\omega)$ corresponds to the actual physical frequency of the emitted light.
The central peak at $\omega \simeq \omega_{\rm cav}$ corresponds to transitions between dressed states that only differ by the number $n$ of cavity photons. The lateral peaks originate from transitions between different dressed states, e.g. at $\omega\simeq 2\Omega_{eg}-\omega_{\rm cav}$ ($|gn\rangle-|en\rangle \rightarrow |g(n+1)\rangle+|e(n+1)\rangle$, solid blue arrows) and $\omega\simeq \omega_{\rm cav}\pm \Omega_{eg}$ ($|fn\rangle \rightarrow |gn\rangle\pm|en\rangle$, dashed and dotted red arrows). Correspondingly to each arrow in the diagram (a), the vertical lines in the spectra (b,c) indicate the expected position of the peak: the agreement with the numerical spectra is excellent.

Similar reasonings can be used to explain the peaks that are apparent in Fig.\ref{fig:scheme_intensity}(b,c) around $\Omega_{eg}=\omega_{\rm cav},\,2\omega_{\rm cav}$: the stronger one at $\Omega_{eg}\simeq 2\omega_{\rm cav}$ originates from the resonant mixing of the $|g0\rangle-|e0\rangle$ and $|f1\rangle$ states by the anti-RWA terms of the emitter-cavity coupling (see the horizontal arrow in Fig.\ref{fig:levels_spectra}(d)). The peaks at $\omega_{\rm cav}\pm \Omega_{\rm cav}/\sqrt{2}$ in the frequency-resolved spectra of Fig.\ref{fig:levels_spectra}(e,f) indeed correspond to transitions to and from the new eigenstates in the form of linear combinations of the $|g0\rangle-|e0\rangle$ and $|f1\rangle$ states.
The interpretation of the weaker and narrower peak at $\Omega_{eg}\simeq \omega_{\rm cav}$ in Fig.\ref{fig:scheme_intensity}(b,c) is similar: the mixing now occurs between the $|g0\rangle-|e0\rangle$ and the $|g2\rangle+|e2\rangle$ states, the weaker matrix element being due to the small non-resonant admixture of $|f1\rangle$ to the $|g2\rangle+|e2\rangle$ state.

The peak around $\Omega_{eg}=2\omega_{\rm cav}$ has a simple physical interpretation in terms of the dynamical Casimir emission: under the effect of the driving laser, the atom performs Rabi oscillations on the $g\leftrightarrow e$ transition, so that the atom-cavity coupling is periodically switched on and off at frequency $2\Omega_{eg}$~\cite{ridolfo}. As usual, amplification of the zero-point fluctuations is most effective when the modulation frequency is close to an even multiple of the cavity frequency~\cite{Lambrecht_rev}. 
The radiated energy is compensated by those $e\rightarrow g$ spontaneous emission processes that move the system upwards in the dressed level scheme of Fig.\ref{fig:levels_spectra}(d), i.e. the spontaneous emission of photons at a physical frequency slightly smaller than the one of the absorbed photons from the drive laser at $\omega_L$.

\begin{figure}[htbp]
\includegraphics[width=0.89\columnwidth,clip]{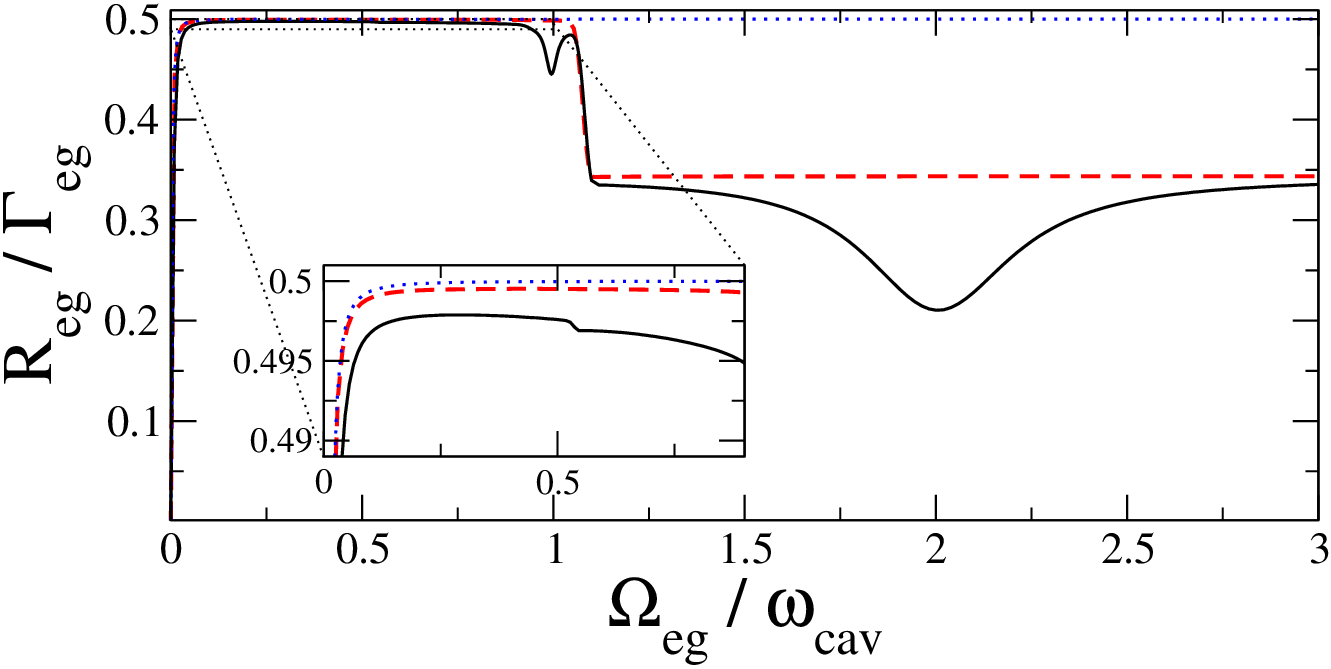}
\caption{(Color online)
Solid line: Rate of photon absorption from the drive laser as a function of its Rabi frequency $\Omega_{eg}$.
Same system parameters as in Fig.\ref{fig:scheme_intensity}. Red dashed line: same curve in the vanishing emitter-cavity coupling case $\Omega_{\rm cav}=0$. Blue dotted line: same curve for a two level emitter.
\label{fig:absorption}}
\end{figure}

We finally turn to the back-action of the emitter onto the drive beam. The simplest effect to consider is the absorption of the drive laser light by the emitter performing the $g\rightarrow e$ transition. Using standard results of quantum optics~\cite{QuantumNoise}, the average rate of photon absorption is related to the imaginary part of the expectation value of the emitter polarization by
$R_{eg}=2\Omega_{eg}\,\textrm{Im}\{\textrm{Tr}[\chd_{eg}\,\rho_{ss}] \}.$
A plot of this quantity as a function of $\Omega_{eg}$ is shown in Fig.\ref{fig:absorption}.
For low $\Omega_{eg}$, the dependence is the standard one of saturated absorption by a two-level atom (blue dotted line). 
The downward jumps at $\Omega_{eg}=\omega_{\rm cav}/2$ and (more visibly) at $\Omega_{eg}=\omega_{\rm cav}$ are closely related to the thresholds previously observed in $I_{\rm cav}$ and $I_{fe}$.

Clear signatures of the back-action effect by the dynamical Casimir emission are visible as absorption dips around $\Omega_{eg}=\omega_{\rm cav}$ and $\Omega_{eg}=2\omega_{\rm cav}$. If the frequency of Rabi oscillations is close to the resonance peak at which dynamical Casimir emission is maximum, the emitter indeed experiences a sizable friction force that tends to oppose the Rabi oscillations and effectively reduces the rate of energy absorption from the drive~\footnote{A similar phenomenon occurs in a resonantly driven harmonic oscillator, where the absorbed energy for a given driving strength scales inversely with the damping rate.}. This interpretation is validated by the observation that both negative peaks completely disappear for $\Omega_{\rm cav}=0$ (red dashed line in Fig.\ref{fig:absorption}). Remarkably, the amplitude of both peaks is a non-negligible fraction of the total absorption rate.

In conclusion, we have theoretically studied the dynamics of an optically driven three-level emitter in a ladder configuration embedded in a single-mode microcavity. This has required taking into account those novel optical processes that stem from the anti-rotating-wave terms of the light-matter coupling Hamiltonian. Rabi oscillations under the effect of a continuous wave driving laser result in a sizeable dynamical Casimir emission from the cavity. A clear signature of the backaction of the dynamical Casimir emission onto the source of the time-modulation is visible as a marked dip in the absorption from the drive beam when the dynamical Casimir emission is the strongest. 
Generalization of this approach to more complex geometries such as optical black holes~\cite{faccio} is expected to shine light on fundamental problems of contemporary physics such as black hole evaporation under the effect of the Hawking radiation~\cite{fabbri}.

IC is grateful to R. Balbinot, A. Fabbri, and S. Savasta for continuous discussions. IC acknowledges financial support from ERC through the QGBE grant.

\end{document}